\documentclass[apj,twocolumn,twocolappendix]{aastex631}
\usepackage{amsmath,amsfonts,amssymb,bm,bbm}
\usepackage[colorlinks=true]{hyperref}
\usepackage{xcolor}
\usepackage{graphbox}
\usepackage{lipsum}
\usepackage{booktabs}

\begin{document}

\title{Insights on Galaxy Evolution from Interpretable Sparse Feature Networks}

\author[0000-0002-5077-881X]{John F. Wu}
\affiliation{Space Telescope Science Institute, 3700 San Martin Dr, Baltimore, MD 21218}
\affiliation{Department of Physics \& Astronomy, Johns Hopkins University, 3400 N Charles St, Baltimore, MD 21218}
\affiliation{Department of Computer Science, Johns Hopkins University, 3400 N Charles St, Baltimore, MD 21218}

\email{jowu@stsci.edu}

\begin{abstract}
Galaxy appearances reveal the physics of how they formed and evolved. Machine learning models can now exploit galaxies' information-rich morphologies to predict physical properties directly from image cutouts. Learning the relationship between pixel-level features and galaxy properties is essential for building a physical understanding of galaxy evolution, but we are still unable to explicate the details of how deep neural networks represent image features. To address this lack of interpretability, we present a novel neural network architecture called a Sparse Feature Network (SFNet). SFNets produce interpretable features that can be linearly combined in order to estimate galaxy properties like optical emission line ratios or gas-phase metallicity. We find that SFNets do not sacrifice accuracy in order to gain interpretability, and that they perform comparably well to cutting-edge models on astronomical machine learning tasks. Our novel approach is valuable for finding physical patterns in large datasets and helping astronomers interpret machine learning results.
\end{abstract}


\section{Introduction}

Galaxies have spectacularly information-rich appearances. For a hundred years, astronomers have categorized galaxies on the basis of morphology and colors seen in optical imaging, and they have noted correlations between galaxies' physical properties and these simple morphological classifications \citep[e.g.,][]{1926ApJ....64..321H,1959HDP....53..373Z,1961hag..book.....S,2008MNRAS.389.1179L,2020IAUS..353..205M}. However, until very recently, there has been no way to connect the physics of galaxy formation and evolution with their \textit{detailed} appearances. 

Deep neural networks that exploit pixel-scale information from images can now robustly predict galaxy properties; optical images of galaxies contain enough information to constrain their physical characteristics \citep[e.g.,][]{2019ApJ...887..251W,2020ApJ...900..142W,2021arXiv211101154B}. In fact, entire galaxy spectra can be estimated directly from image cutouts of individual galaxies \citep{2020arXiv200912318W,2021ApJ...914..142H,Doorenbos+2022,Doorenbos+2024,2024MNRAS.531.4990P}. This last discovery highlights the promise of machine learning (ML) in the physical sciences. By leveraging deep learning models such as convolutional neural networks (CNNs), we can derive valuable physical constraints from plentiful galaxy images \citep{2023PASA...40....1H,2024MNRAS.531.4990P}.
So \textit{machines} are capable of learning physics from galaxy imaging---but what about us \textit{astronomers}?

Two major challenges arise when we try to interpret physics from deep neural networks. First, each neuron in the network represents a non-linear combination of features, resulting in uninterpretable mess of features \citep[known as the ``superposition'' problem;][]{superposition}. Second, each neuron can also be activated in multiple independent contexts, and can thus take on multiple meanings \citep[known as ``polysemanticity,'' see e.g.,][]{geva2021transformerfeedforwardlayerskeyvalue}. In other words, neuron values are not one-to-one with learned concepts. Superposition and polysemanticity allow neural networks to increase their capacity for storing features \citep{olah2020zoom,superposition,10136140}; these phenomena are topics of cutting-edge ML research, and we present an extended discussion in Appendix~\ref{appendix}. Unfortunately, they also obscure the interpretation and contribution of any specific neuron, preventing us from understanding the physics behind the patterns that ML models learn.

Fortunately, there exists a strategy to combat these issues in ML. Sparse coding represents data using a small number of active elements from a larger set of basis functions \citep{Olshausen1996,NIPS2006_2d71b2ae}, making it easier to inspect the individual features that contribute to ML model predictions \citep[e.g.,][]{olah2017feature}. Recently, sparse ML algorithms have been used in the artificial intelligence community to disentangle the internal activations of large language models \citep{radford2019language}, and mitigate the problems of superposition and polysemanticity \citep{makhzani2014ksparseautoencoders,cunningham2023sparseautoencodershighlyinterpretable,2024arXiv240800657O,templeton2024scaling}. Therefore, sparse coding can reveal the mechanics of how deep neural networks learn. 

We introduce a novel, interpretable CNN for images: a Sparse Feature Network (SFNet).
CNNs encode image data from a high-dimensional pixel space ($n \sim 10^5$ pixels) into a lower-dimensional latent space ($d \sim 10^3$ features). 
To disentangle the learned features, we enforce a $k$-sparse constraint while training our ML model: only $k \ll d$ features are allowed to have non-zero activations for each image example. By keeping just the top $k$ activating features, sparse algorithms are forced to consolidate features into the fewest possible neurons.
Our SFNet activates $k$ features per image in its penultimate layer, and makes predictions by using a linear combination of the sparse activations. 
Since we only add a top-$k$ constraint to a standard CNN architecture, this modification incurs zero extra computational cost or memory.

We train the SFNet to predict galaxy properties from image cutouts according to the methodology described in Section~\ref{sec:methodology}. In Section~\ref{sec:results}, we show that the SFNet achieves robust performance while learning interpretable features. We discuss these results further in Section~\ref{sec:discussion}, and state our conclusions in Section~\ref{sec:summary}. 

\begin{figure*}[ht]
    \centering
    \includegraphics[width=\linewidth]{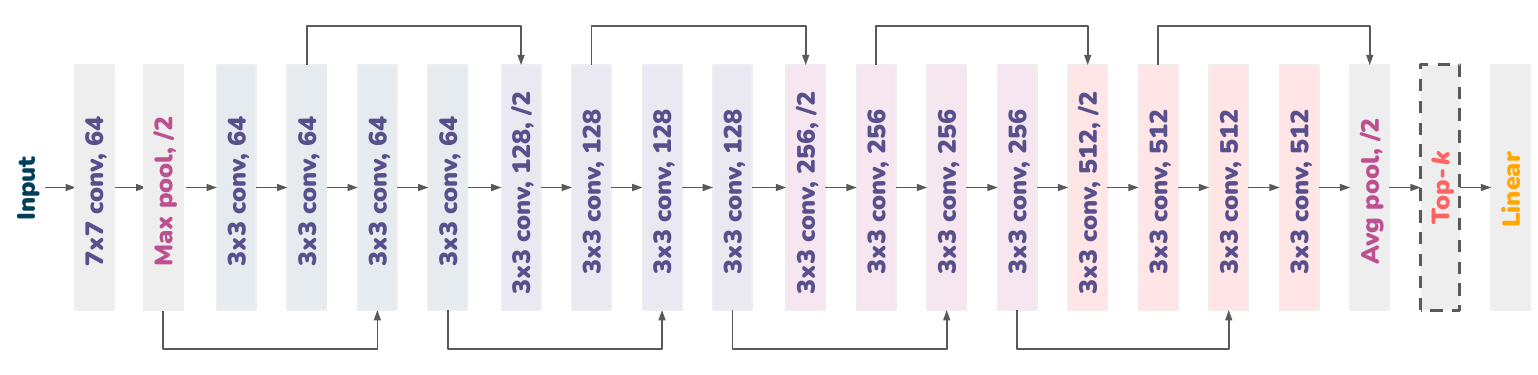}
    \caption{{\bf Our SFNet architecture, which resembles a resnet18 modified with a penultimate ``Top-k'' layer to ensure interpretable, sparse image features.} A \texttt{3x3 conv, 128, /2} block denotes a $3\times 3$ 2D convolution layer with 128 channels that downsamples the image size by a factor of 2. Each convolution layer is followed by batch normalization and then a ReLU activation function. Arrows show how inputs flow through the network, and when two inputs are sent as inputs to the same layer, they are concatenated together (i.e. a residual connection). %
    }
    \label{fig:SFNet-architecture}
\end{figure*}

\section{Methodology} \label{sec:methodology}

We select galaxies from the SDSS Main Galaxy Sample \citep{2002AJ....124.1810S,sdssdr7} that have emission line signal-to-noise ratios greater than 3 for [\ion{N}{2}] $\lambda6584$, H$\alpha$ $\lambda6564$, [\ion{O}{3}] $\lambda5007$, and H$\beta$ $\lambda4861$, where each line is designated by its rest wavelength in Angstroms. In addition to galaxy line fluxes, we compile physical properties derived from fitting spectral energy distribution models to SDSS photometry and spectra \citep{2003MNRAS.346.1055K,2004MNRAS.351.1151B,2004ApJ...613..898T}. For all galaxies, we obtain $gri$-band $144\times 144$ image cutouts resampled at $0.262$ arcsec per pixel resolution from the Legacy Survey website \citep{2019AJ....157..168D}. Our SDSS sample comprises 250,207 galaxies in total.

Our SFNet implementation is shown in Figure~\ref{fig:SFNet-architecture}. We use a CNN architecture identical to the resnet18 \citep{he2015deepresiduallearningimage}, except that we insert a top-$k$ operation before the final linear layer. 
This additional top-$k$ layer guarantees that each prediction is a linear combination of $k$ sparse features.
We use $d=512$ latent features (after the average pooling layer) and set $k=4$. Later, in Section~\ref{ssec:comparison-pca}, we vary $k \in \{2, 3, 4, 6, 8\}$. Thus, the SFNet learns a representation of each galaxy using just a few features out of a library of 512 possible features.
Critically, the features activate directly after the convolution layers, so that morphological features are localized in pixel space.

The resnet18 backbone of the model is initialized to ImageNet-pretrained weights \citep{5206848}, and we update model parameters by minimizing the root mean squared error. We randomly designate 20\% of objects for validation, and train on the remainder. Random flips are applied for data augmentation. We train for 20 epochs using the ranger optimizer \citep{wright2021ranger21synergisticdeeplearning} and an initial learning rate of 0.1 with a flat+cosine annealing schedule. All code and the trained models presented here can be found in our public Github repository: \url{https://github.com/jwuphysics/sparse-feature-networks}.

\section{Results} \label{sec:results}

Motivated by previous works that connect galaxy imaging with their physical properties \citep{2019MNRAS.484.4683W,2020arXiv200912318W,2021MNRAS.506.3313G,2021ApJ...914..142H}, we train the SFNet to predict spectroscopic line fluxes (Section~\ref{ssec:ionization}) and gas-phase metallicity (\ref{ssec:metallicity}) directly from image cutouts.
We identify the learned features using their sparse activation index, e.g., \texttt{SL\,17} for the 17th feature of a SFNet trained to predict spectral lines fluxes, or \texttt{Z\,256} for the 256th feature of a SFNet trained to predict gas-phase metallicity. Feature activations are scaled between 0 and 1, and we refer to these example activation strengths as $A_{\tt SL\,17}$ and $A_{\tt Z\,256}$, respectively.  %

\subsection{Emission line fluxes} \label{ssec:ionization}

We train a SFNet to predict the spectral line fluxes for [\ion{N}{2}], H$\alpha$, [\ion{O}{3}], and H$\beta$. These lines can diagnose a galaxy's global ionization state; for example, the line ratios [\ion{N}{2}]/H$\alpha$ against [\ion{O}{3}]/H$\beta$ comprise the BPT diagram \citep{1981PASP...93....5B}, which is commonly used to differentiate activate galactic nuclei (AGN) from star-forming galaxies. Other physical processes can also affect these line strengths, including stellar evolution/age, AGN, gas metallicity, gas physical state, and dust attenuation.

The eight most frequently activated features explain 99.99\% of the variance in the dataset, but there is significant correlation between them. The four most frequently activated features can still explain 91.6\% of the variance; therefore, we restrict our investigation to these top four activations that correspond to features \texttt{SL\,17}, \texttt{SL\,138}, \texttt{SL\,157}, and \texttt{SL\,322}. %

\begin{figure*}[ht]
    \centering
    \includegraphics[width=0.49\textwidth]{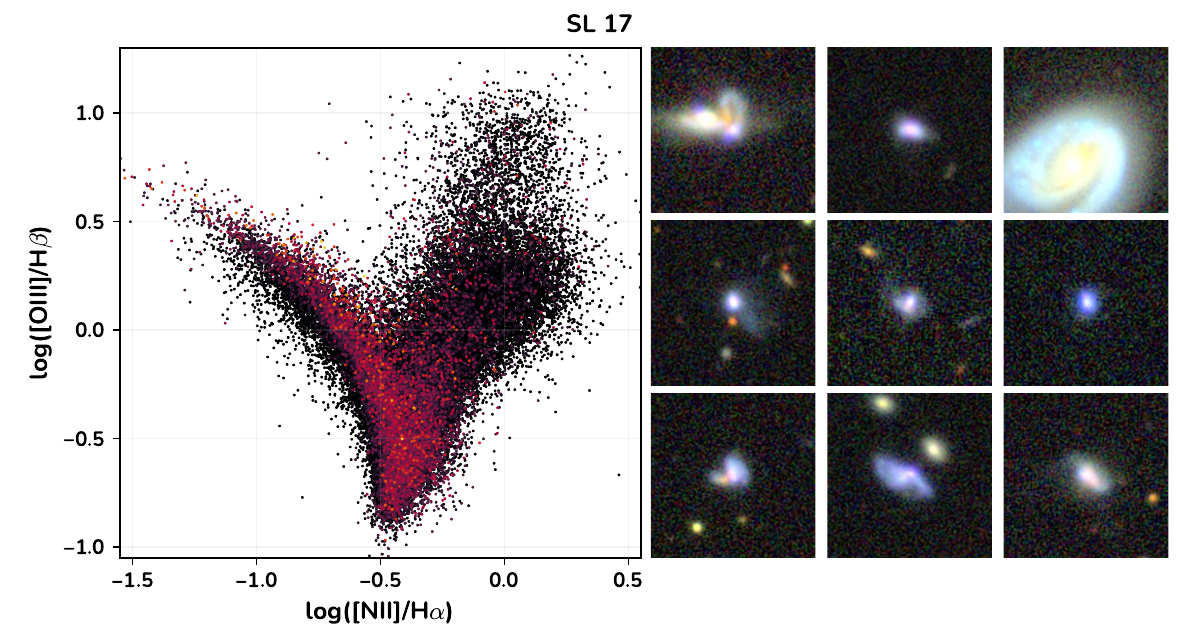}
    \includegraphics[width=0.49\textwidth]{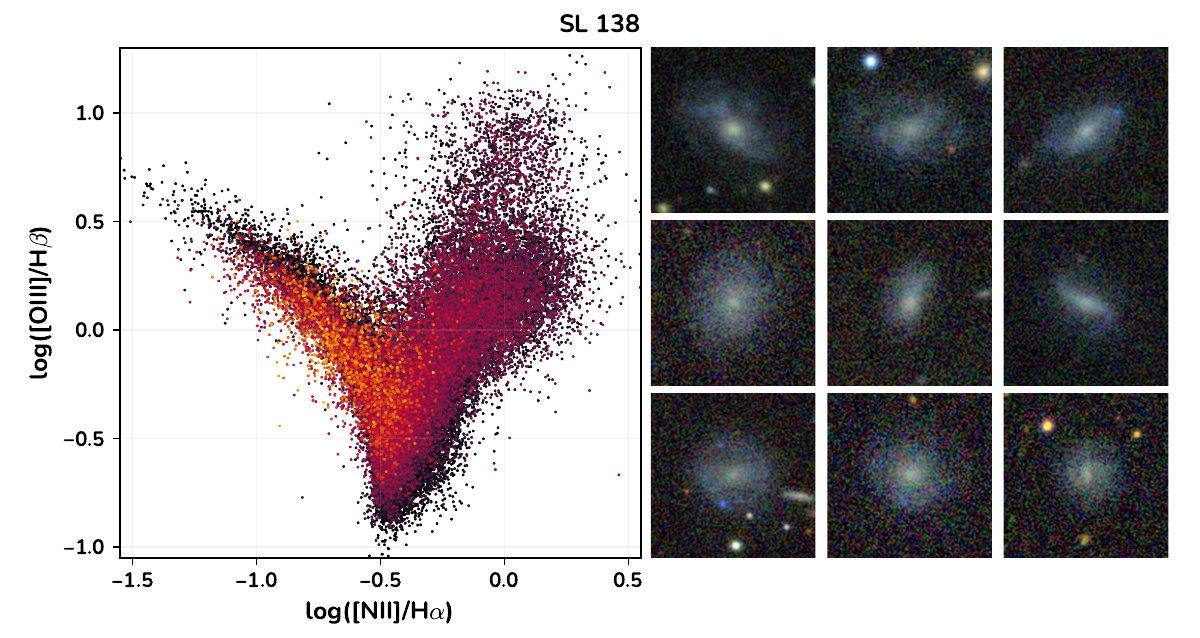}
    \includegraphics[width=0.49\textwidth]{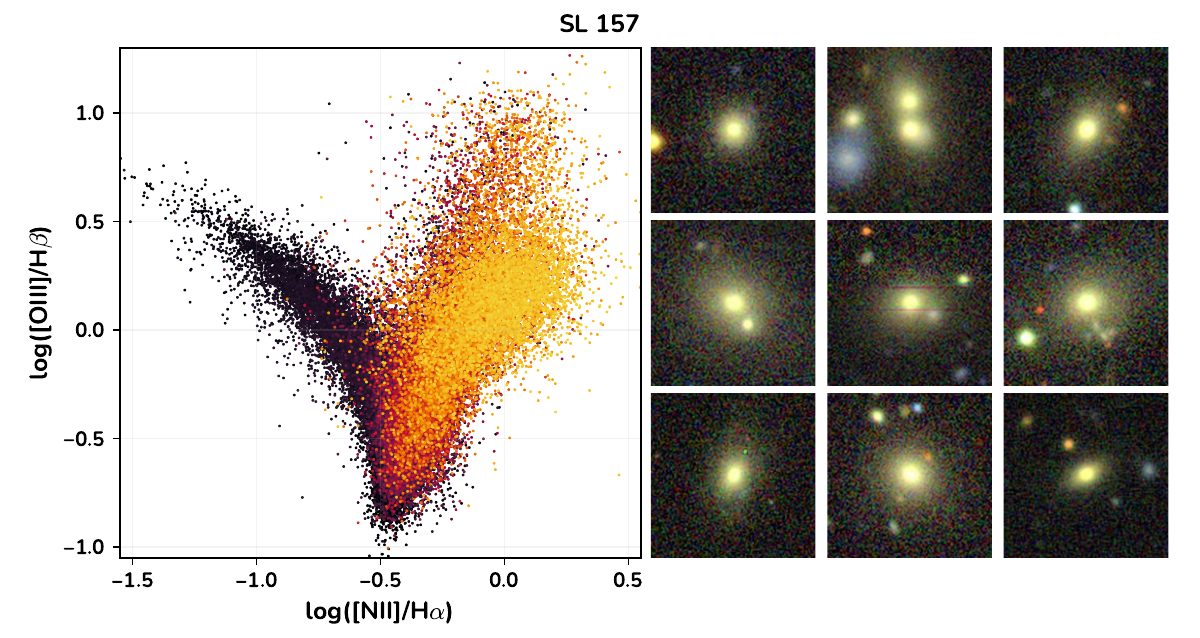}
    \includegraphics[width=0.49\textwidth]{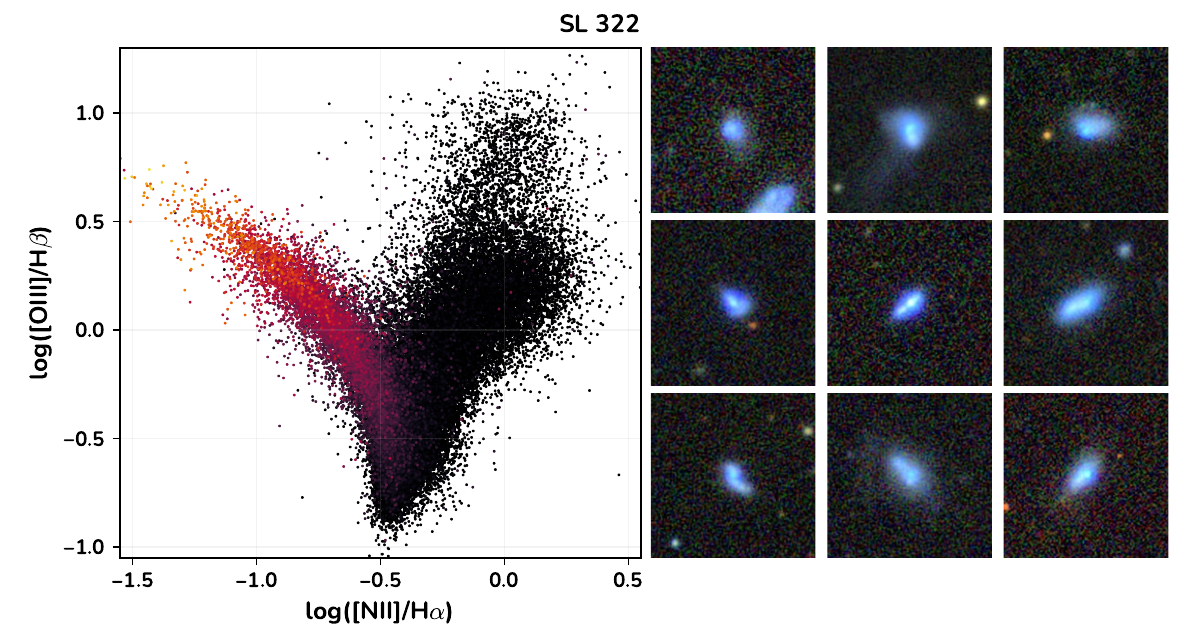} 
    \caption{\textbf{SFNet learned features when trained to predict optical emission lines.} We show how each feature correlates with optical line ratios (BPT diagrams; \textit{left}) and examples of the top nine image activations (\textit{right}). For the BPT diagrams, colors denote scaled activation strength, ranging from black (zero) to dark magenta (low) to bright yellow (high).}
    \label{fig:bpt-interpretation}
\end{figure*}

In Figure~\ref{fig:bpt-interpretation}, we show galaxy image examples for the strongest activations, and where they appear in a BPT diagram. The four features reside in overlapping regions of the [\ion{O}{3}]/H$\beta$ versus [\ion{N}{2}]/H$\alpha$ space, but they all peak in distinct locations. For example, the locus of points for \texttt{SL\,17} appear to be at higher [\ion{N}{2}]/H$\alpha$ and [\ion{O}{3}]/H$\beta$, indicating harder ionizing radiation and higher gas density \citep[based on nebular diagnostics, e.g.,][]{2019ARA&A..57..511K}. Meanwhile, the opposite appears to be true for \texttt{SL\,138} (or at least, for non-AGN with high $A_{\tt SL\,138}$). Additional features contain information about metallicity and stellar populations. In Table~\ref{tab:features}, we offer physical interpretations for these features. %

\begin{deluxetable*}{lcll}[t!]
\tablecolumns{4}
\tablecaption{Image features learned by a SFNet to independently predict optical spectral line fluxes and metallicity. We only show features that are most frequently activated. \label{tab:features}
}
\tablehead{
    \colhead{Feature} & 
    \colhead{$f_{\rm activated}$} &
    \colhead{Morphology} &
    \colhead{Physical interpretation}
}
\startdata
\texttt{SL\,17} & 62.9\% & very blue, compact or merging & extreme star formation, hard ionizing radiation \\
\texttt{SL\,138} & 87.0\% & low surface brightness, face-on & low gas density, soft ionizing radiation \\
\texttt{SL\,157} & 95.8\% & red, elliptical & old stellar populations, high mass, high metallicity \\
\texttt{SL\,322} & 62.9\% & blue, irregular & low mass, low metallicity, star-forming \\
\hline
\texttt{Z\,61} & 99.6\% & red, face-on, bright core & high metallicity \\
\texttt{Z\,256} & 87.9\% & blue, edge-on disks & low metallicity \\
\enddata
\end{deluxetable*}

\subsection{Gas metallicity} \label{ssec:metallicity}

We separately train a SFNet to predict the gas-phase metallicity, $Z_{\rm gas} = 12 + $log(O/H), directly from images. We remove all objects with invalid estimates for metallicity, stellar mass, and star formation rate, leaving 117,223 galaxies from the SDSS Main Galaxy Sample. The scatter of metallicity is $0.207$~dex, and our trained SFNet can predict it to within $0.087$~dex using only two image features, \texttt{Z\,61} and \texttt{Z\,256}, which are activated 99.6\% and 87.9\% of the time respectively. If we also incorporate the third and fourth most frequent activations (40.2\% and 11.9\%, respectively), then we can achieve $0.086$~dex error. This level of performance suggests that our SFNet is extremely accurate (see Section~\ref{ssec:performance-interpretability} for further discussion).

In the bottom two rows of Table~\ref{tab:features}, we describe the two main features used to predict metallicity. \texttt{Z\,61} is associated with galaxies with prominent red bulges and is linked to higher metallicity. It activates strongly for red ellipticals or face-on disk galaxies (often with two loosely wound spiral arms; see, e.g., \citealt{2019MNRAS.487.1808M} for more on the correlations between galaxy properties and spiral arm pitch angle).  Feature \texttt{Z\,256} is associated with blue, edge-on disk galaxies and lower metallicity. 
\citet{2004ApJ...613..898T} noted that metal-poor galaxies tended to have highly inclined disk morphologies, and attributed this effect to observational bias. SDSS spectroscopic fibers collect light from lower-$Z_{\rm gas}$  outskirts of disk galaxies observed edge-on, but would miss those regions for the same galaxies observed face-on. Our SFNet now captures this same observational bias, affecting the library of morphological features that it learns.

\section{Discussion}\label{sec:discussion}

\subsection{Physical laws from galaxy images} \label{ssec:physics}

\begin{figure*}
    \centering\includegraphics[width=0.95\textwidth]{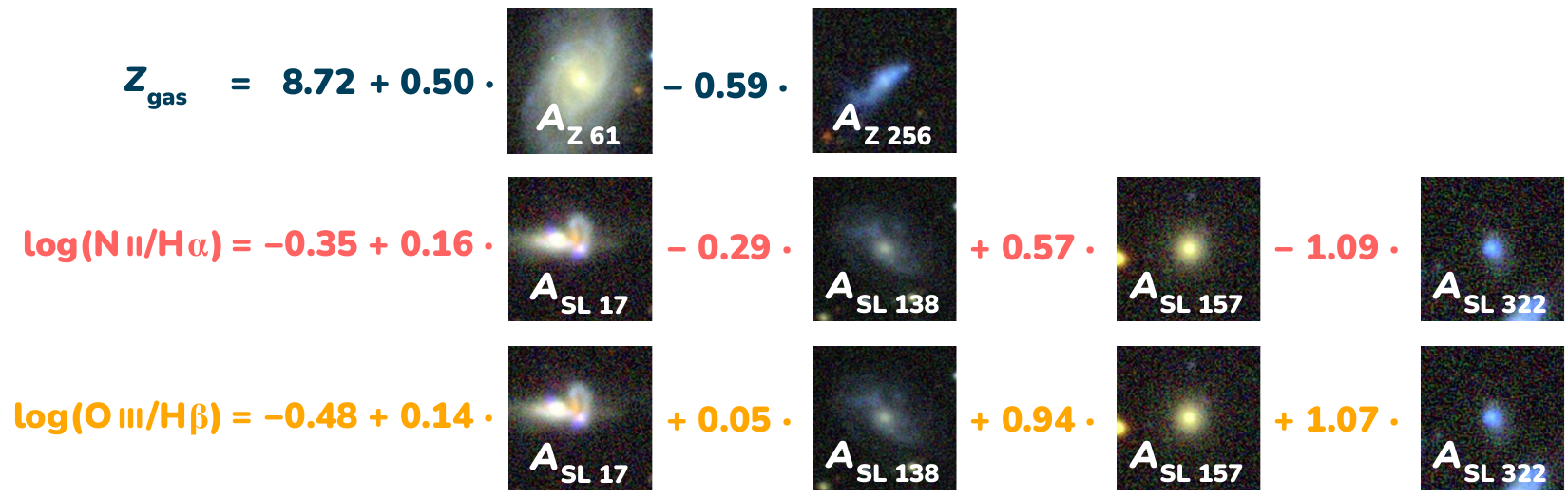}
    \caption{\textbf{Linear relationships between galaxy properties and image features learned by the SFNet.} We show how the gas metallicity, [\ion{N}{2}]/H$\alpha$, and [\ion{O}{3}]/H$\beta$ depend on SFNet feature activations, where SFNet features are represented using the top-activating galaxy image cutout from the validation set.  \label{fig:equations}}
\end{figure*}

Because the SFNet learns a linear projection from these activations to the final prediction, we can write down a simple ``equation'' between learned galaxy features and the physical quantity of interest. In Figure~\ref{fig:equations}, we present linear equations for predicting gas metallicity and spectral line ratios by using the aforementioned galaxy image features.
The maximally activating galaxy examples are shown in the equations (each labeled by their activated feature index).

These linear relations are so simple that we can directly interpret them. The \texttt{Z\,61} and \texttt{Z\,256} features used to predict $Z_{\rm gas}$ directly correspond to higher and lower metallicity. Interestingly, both can be activated at the same time---and in fact, both are non-zero for the majority of galaxies in the validation set, demonstrating the physical interplay between these morphological features and metallicity. For predicting spectral lines, features \texttt{SL\,17} and \texttt{SL\,157} positively correlate with both [\ion{N}{2}]/H$\alpha$ and [\ion{O}{3}]/H$\beta$; we might interpret these morphological features as indicators of spectral hardness and low ionization (nuclear) emission-line regions, respectively. Meanwhile, \texttt{S\,322} correlates with [\ion{O}{3}]/H$\beta$ but anticorrelates with [\ion{N}{2}]/H$\alpha$, suggesting that it is associated with harder ionizing spectra.

The model is only able to learn physical laws by leveraging galaxy morphology present within the optical image cutouts. However, galaxy evolution is governed by the complex interplay between multiple physical processes \citep[e.g.,][]{2015ARA&A..53...51S,2017MNRAS.471.2687B}, which are not always imprinted in galaxy morphology, or perhaps not at the scales or wavelengths probed by our imaging surveys. Nonetheless, CNNs are surprisingly powerful at predicting the spectroscopic tracers of galaxy evolution. \cite{2019A&A...624A.102D} and \cite{2019MNRAS.484.4683W} showed that galaxies' stellar masses can be estimated directly from image cutouts, and \cite{2021MNRAS.501.4359Z} suggest that CNNs are sensitive to small-scale morphological perturbations due to dust in galaxies. Even information outside of the image cutout can be indirectly probed; \cite{2020ApJ...900..142W} measure impact of galaxy environment on CNN-predicted gas properties, suggesting that the well-known density-morphology relation \citep{1980ApJ...236..351D} can be implicitly learned via ML models.
In summary, SFNets can learn a limited (but still powerful) set of physical relationships that govern galaxy evolution.

\subsection{Performance versus interpretability} \label{ssec:performance-interpretability}

Previous works have found that a CNN trained on optical image cutouts can successfully classify AGN from star-forming galaxies with an accuracy \citep[e.g.,][]{2021ApJ...914..142H,2022arXiv221207881G,Doorenbos+2024}. We test how well the interpretable SFNet from Section~\ref{ssec:ionization} can also identify AGN. By optimizing a linear support vector machine (SVM) classifier on \texttt{SL\,17}, \texttt{SL\,138}, \texttt{SL\,157}, and \texttt{SL\,322} from the training dataset, we can classify AGN with a validation accuracy of 0.85 and F1 score of 0.72.\footnote{If we use all non-zero SFNet features, then the best-fit linear SVM validates at 0.83 accuracy and 0.70 F1 score, i.e., it performs \textit{worse}. This suggests that restricting to the four most frequent feature activations reduces noise---another benefit of sparsity.} In comparison, the highly tuned CNN from \cite{2022arXiv221207881G} achieves an accuracy of 0.89 and F1 score of 0.75, and the conditional diffusion model by  \cite{Doorenbos+2024} results in an accuracy of 0.82 and F1 score of 0.73 (although we note that they use a slightly different AGN classification scheme). Our experiments demonstrate that a \textit{linear combination} of our interpretable features can distinguish AGN from star-forming galaxies at a level comparable to state-of-the-art ML algorithms.

In Section~\ref{ssec:metallicity}, we estimate gas metallicity to within 0.087~dex using a linear combination of two features (and 0.086~dex using four features).
For comparison, a typical CNN only achieves $0.085$~dex error \citep{2019MNRAS.484.4683W}; note that the systematic scatter is $\sim 0.03$~dex. Again, this suggests that SFNets do not sacrifice performance in order to gain interpretability.

\subsection{Comparison to PCA} \label{ssec:comparison-pca}

While the SFNet produces interpretable features by imposing a sparsity constraint during the training process, other techniques are frequently used to reduce the number of learned features.
Here we compare SFNet features against another common approach: dimensionality reduction of learned model features via principal components analysis (PCA). 
Given feature activations in the penultimate layer of a regularly trained CNN, we can obtain the principal components by computing eigenvectors of the dense activation's covariance matrix. They are listed in descending order of how much variance is accounted for by each principal component (i.e., ordered by descending eigenvalues).
By selecting the top $k$ principal components, we can project dense feature activations into a smaller space of sparse activations.

\begin{figure*}
    \centering 
    \includegraphics[width=0.95\textwidth]{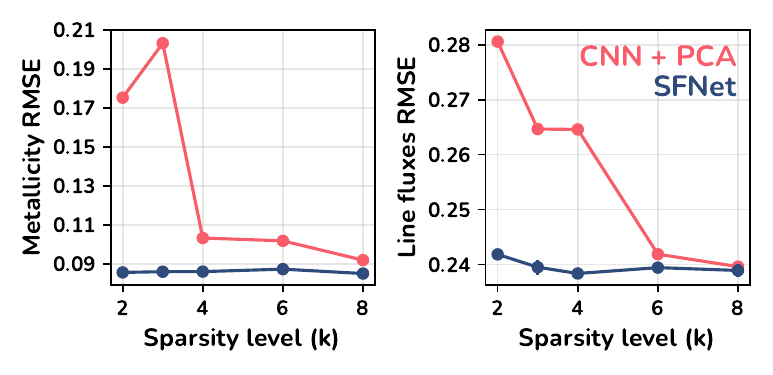}
    \caption{\textbf{SFNet comparison against a CNN with dimensionality-reduced features.} Using the SFNet model (blue) and a dense CNN with top principal components as features (red), we report RMSE values (lower is better) for models trained to predict metallicity (\textit{left}) and spectral line fluxes (\textit{right}) for several values of $k$ (i.e., sparsity or number of PCA components).
    \label{fig:pca-comparison}}
\end{figure*}

We test whether this smaller set of principal components are as robust as the SFNet features.
We train a multivariate linear regression model using the first $k$ principal components from dense CNN activations as features, again using the same training and validation split, to predict both the metallicity and spectral line strengths.
For comparison, we have also re-trained the SFNet using $k \in \{2, 3, 4, 6, 8\}$. We independently re-train the SFNet three times in order to estimate the mean and standard error on the prediction RMSE.\footnote{We use the same training/validation split and re-initialize the model each time for each value of $k$. Because we repeat the training process from scratch, the RMSE values presented in this section slightly differ from the ones reported in Section~\ref{sec:results}.}
The results are shown in Figure~\ref{fig:pca-comparison}, and specific RMSE values are given in Table~\ref{tab:dimred}.

\begin{deluxetable}{l cc cc}[t!]
\tablecolumns{5}
\tablecaption{Comparison of performance using a SFNet and a CNN with PCA-based dimensionality reduction. \label{tab:dimred}
}
\tablehead{
    \colhead{} &
    \multicolumn{2}{c}{RMSE: Metallicity} &
    \multicolumn{2}{c}{RMSE: Spectral Lines}
    \\ 
    \cmidrule(lr){2-3}\cmidrule(lr){4-5} 
    \colhead{$k$} & 
    \colhead{SFNet} &
    \colhead{PCA} &
    \colhead{SFNet} &
    \colhead{PCA}
}
\startdata
2 & $0.0858 \pm 0.0004$ & 0.1753 & $0.2418 \pm 0.0002$ & 0.2806 \\
3 & $0.0862 \pm 0.0011$ & 0.2032 & $0.2395 \pm 0.0014$ & 0.2646 \\
4 & $0.0862 \pm 0.0013$ & 0.1034 & $0.2383 \pm 0.0003$ & 0.2646 \\
6 & $0.0875 \pm 0.0007$ & 0.1020 & $0.2394 \pm 0.0009$ & 0.2419 \\
8 & $0.0852 \pm 0.0004$ & 0.0921 & $0.2389 \pm 0.0012$ & 0.2396 \\
\enddata
\end{deluxetable}

For both metallicity and spectral line fluxes, $>95\%$ of the variance is explained using the top $k=8$ principal components.
For metallicity (left panel of Figure~\ref{fig:pca-comparison}), we find that the CNN + PCA performs poorly on the validation set for all values of $k$. Meanwhile, the SFNet can reconstruct metallicity with only $k=2$ explainable features, reinforcing our results found in Section~\ref{ssec:metallicity}. 
We interpret this as evidence that the metallicity cannot be linearly combined using a small number of principal components of features extracted via a typical CNN.\footnote{The non-monotonic decrease in RMSE with increasing $k$ suggests that PCA---while able to reduce dimensionality---does not optimally distill morphological features into the ones most relevant for predicting galaxy properties.}
We turn our attention to spectral line fluxes (right panel of Figure~\ref{fig:pca-comparison}), and find that performance is poor for small values of $k$, but if we use 8 principal components, then the RMSE becomes comparable to that of the SFNet.
The SFNet can accurately estimate spectral line fluxes using only $k=4$ features (with significantly worse performance at $k=2$).

Our experiment shows that we can reduce the dimensionality of CNN feature activations via PCA, but the ensuing principal components are not always robust features for recovering the training targets (e.g., for metallicity).
These difficulties may arise from superposition and polysemanticity, which cause the most important features to be entangled and spread across multiple CNN activations in highly non-linear ways.
Since typical CNNs are not optimized with any sparsity constraint in mind, there is no guarantee that a linear combination of the learned features should be interpretable (although there are non-linear methods for finding optimal summary statistics; e.g., \citealt{Charnock_2018}).
By forcing a CNN to be sparse during the training process, we find that SFNets can learn robust features and produce accurate predictions.

\subsection{Comparison to GalaxyZoo features}

Do regular CNNs also learn interpretable features?
We compare the SFNet against Zoobot, a CNN trained to classify GalaxyZoo morphologies \citep{2008MNRAS.389.1179L,2023JOSS....8.5312W,Walmsley+2024}. We consider the resnet18 variant of Zoobot, which has the same architecture as our SFNet modulo the top-$k$ sparsity constraint (Figure~\ref{fig:SFNet-architecture}). Zoobot predicts citizen scientist vote fractions for GalaxyZoo morphological labels. We hypothesize that Zoobot's 512-dimensional features directly after the pooling layer suffer from superposition and polysemanticity, and thereby produce an uninterpretable combination of morphology and color.  

Using morphological features from Zoobot, we train a regression model to predict galaxy metallicity. This experiment allows us to assess whether traditional CNNs can also be used to learn astrophysical relationships between image features and galaxy properties.  We use the same training/validation datasets from Section~\ref{ssec:metallicity}, and independently test two models that use Zoobot-extracted features as inputs and $Z_{\rm gas}$ as training targets. 

First, we fit a linear model to the training dataset, and find poor performance on the validation dataset (error of $0.183$~dex). When we fit on the \textit{validation} dataset---i.e., if we attempt to memorize the answers---the error remains very high ($0.181$~dex). This result demonstrates that no linear combination of Zoobot features can reconstruct galaxy metallicity.

Second, we fit an gradient-boosted decision tree (XGBoost) model to the training dataset, and achieve only $0.180$~dex error on the validation dataset. However, when we ``cheat'' by fitting on the validation dataset, we can lower the error down to $0.085$~dex (comparable to the SFNet performance). 

Our experiments confirm that the Zoobot features indeed contain enough information to memorize the validation set, but that these features (\textit{i}) cannot be linearly combined to reconstruct galaxy metallicity, and (\textit{ii}) cannot generalize across the random training/validation split. We surmise that regular CNN features are not interpretable. Meanwhile, our SFNet is both performant and interpretable.

Finally, we check whether the SFNet can capture information not found within the 512-dimensional Zoobot feature vector. We optimize a ridge regression model to predict SFNet feature values  using the Zoobot features, finding $R^2$ values of $0.253$ and $0.107$ for the two primary SFNet features, \texttt{Z\,61} and \texttt{Z\,256}, respectively. These low coefficients of determination suggest that our SFNet extracts different morphological features than Zoobot.

\subsection{Limitations of previous interpretability methods}

The ML discipline generally focuses on classification problems, which has led to rapid development of classification-based interpretability algorithms \citep[e.g.,][]{erhan2009_visualizing,zeiler2013visualizingunderstandingconvolutionalnetworks,simonyan2014deepinsideconvolutionalnetworks,8099837}. For example, Gradient-weighted Class Activation Mapping (GradCAM; \citealt{gradcam}) enables researchers to interrogate the model to reveal the pixels that are most important for each class prediction. However, GradCAM and its variations are reliant on classification labels.
For example, galaxy morphologies and other properties exist along a continuum rather than in distinct categories. Even identifying AGN via the BPT diagram, which we explore in Section~\ref{ssec:ionization}, is more accurately posed as a regression problem than a classification problem.\footnote{While there is promise in improving the semantic content of some classification labels \citep[e.g.,][]{2022arXiv221014760B}, categorization is often an oversimplification of richer (astro)physics.} 

Moreover, interpretation for classification tasks (with fully separable classes) serves a different purpose than interpretation for regression tasks. Specifically, astronomers might ask questions like, ``which morphological features in this galaxy scale with its gas-phase metallicity?'' Astronomers are generally less interested in asking questions like, ``which morphological features in this image scale with the probability that it is a galaxy rather than a star?'' %
Regression tasks can be interpreted by understanding how output quantities scale with input features (see, e.g., our ``equations'' in Figure~\ref{fig:equations}), but classification problems---wherein predictions are selected via the softmax operation---only require the dominant class to be marginally higher than every other class. 
Therefore, interpretability for regression and classification problems is different.

Other interpretability methods like saliency mapping \citep{simonyan2014deepinsideconvolutionalnetworks} cannot learn the non-linear patterns of multiple pixels at once---which is exactly what makes deep learning models so powerful. 
Despite these limitations, saliency and GradCAM-like techniques are widely adopted in astronomy \citep[e.g.,][]{2019ApJ...882L..12P,2020ApJ...900..142W,2021MNRAS.501.4579B,2024ApJ...967..152A}. 
Our results demonstrate that sparse algorithms provide a powerful alternative for interpretable ML in the physical sciences.

\section{Conclusions} \label{sec:summary}

We present a novel method for learning the relationship between the physical properties and interpretable morphologies of galaxies. Our sparse feature network (SFNet; Figure~\ref{fig:SFNet-architecture}) maps galaxy features onto galaxies' physical properties such as spectroscopic line fluxes and gas-phase metallicity. We identify features that correspond to hard ionizing radiation (starbursts), low gas density, old stellar populations/high metallicities, and star-forming regions/low metallicity (e.g., Table~\ref{tab:features} and Figure~\ref{fig:bpt-interpretation}). Critically, our method does not substantially sacrifice ML performance in order to gain interpretability. We can write down ``equations'' (Figure~\ref{fig:equations}) that reveal the robust linear connection between galaxy properties and their appearances. 
Optimizing a SFNet yields more robust and interpretable features than training a CNN and then using PCA to consolidate the dense features: we find that a linear combination of $k$ interpretable morphological features produces more accurate predictions than by using the top $k$ principal components of a regular CNN's features (see Figure~\ref{fig:pca-comparison}).

We do \textit{not} imply that astrophysics can or should be automated by machines. As we discuss in Section \ref{sec:discussion}, astronomers are still necessary for interpreting image features. While the SFNet can highlight what ML algorithms are capable of learning, they cannot replace domain experts, who are essential for scrutinizing the physical interplay between learned features and predicted quantities.

This work represents important progress toward developing explainable deep learning algorithms for the physical sciences. While we find strong results by incorporating a simple top-$k$ sparsity layer into a resnet18, we have not comprehensively varied the optimization procedure or model hyperparameters. However, we have tested a few values of $k$ and found that the SFNet performance is robust. Additional sparsity constraints, such as decorrelation layers (e.g. zero-phase component analysis or Cholesky whitening) or regularization terms (e.g., L1 norm) in the loss function, may encourage the SFNet to learn improved features. Using a CNN architecture with a smaller receptive field (i.e., by eschewing pooling layers), which guarantees that learned features are localized to a small number of pixels, may also enhance interpretability of SFNets. To aid the physical interpretation of features, we could also deploy an activation maximization model to specifically hone in on activating features \citep[e.g.,][]{olah2017feature}.

\subparagraph{Acknowledgments.}
We thank Christian Jespersen, Charlie O'Neill, Joshua Peek, and Christine Ye for useful conversations.

\appendix

\section{Why are deep neural networks so hard to decipher?}\label{appendix}

\subsection{Superposition and polysemanticity}

Modern deep neural networks are composed of many layers, each with many neurons. 
AI and ML researchers are actively seeking to understand the semantic meanings captured by these neurons.
Mechanistic interpretability is a subfield within AI/ML that aims to clarify the mechanisms by which deep neural networks learn \citep[e.g.,][]{olah2020zoom,rai2024practicalreviewmechanisticinterpretability}, usually in simplified settings (since studying all neurons in all layers at once is infeasible).
Given the rise of AI/ML techniques in the physical sciences, it is also vital to understand how and what our trained neural networks have learned.

However, it is challenging to decipher what features are represented by specific neurons, primarily due to two phenomena: superposition and polysemanticity \citep{superposition}. \begin{itemize}
    \item \textit{Superposition} is where semantic concepts (i.e., features) become distributed across many neurons. For example, a single feature like ``three-armed spiral galaxy'' can be spread out across multiple neurons (in one or even multiple layers). %
    \item \textit{Polysemanticity} means that any given neuron can also have multiple separate meanings. These semantic concepts may not even be correlated with each other! For example, a specific neuron in a regularly trained CNN might activate on features such as ``three-armed spiral galaxy'' and ``$r$-band light bleed from nearby saturated star.''
\end{itemize}
So rather than a one-to-one mapping of neurons to features, or vice versa, there is actually a many-to-many mapping. Multiple learned features can be distributed and superposed across many different neurons. Superposition and polysemanticity encapsulate why it is so challenging to understand neural networks, even in just a single layer.

\subsection{Sparse autoencoders and SFNets}

Originally motivated by biological vision systems \citep{OLSHAUSEN19973311,NIPS2007_4daa3db3}, sparse dictionary learning has emerged as a viable method for understanding (and consolidating) learned features into a small number of neurons. 
Algorithms such as sparse autoencoders \citep[SAEs;][]{makhzani2014ksparseautoencoders,cunningham2023sparseautoencodershighlyinterpretable} are capable of explaining learned features in large language models \citep{bricken2023monosemanticity,templeton2024scaling,2024arXiv240800657O} and deep convolutional neural networks \citep{gorton2024missingcurvedetectorsinceptionv1}.
These recent works have demonstrated that sparse algorithms can disentangle the features represented by deep neural networks.

Previous works with SAEs have typically operated on large, pretrained models and sought to extract the meanings of learned activations \citep[e.g.,][]{bricken2023monosemanticity}.
SAEs are tasked with reconstructing the uninterpretable mess of neuron activations from a specific network layer by learning a set of new sparse activations.
However, SAEs only have two layers, which limits their ability to accurately distill the dense network activations into a few sparse ones.

Our work builds on insights from the mechanistic interpretability community.
Instead of tasking a SAE to interpret learned features from a trained deep CNN, we devise a neural network architecture that is not only expressive like a regular CNN, but also employs sparse dictionary learning to produce interpretable features.

\bibliography{main} %
\bibliographystyle{aasjournal}

\end{document}